# Stabilized Hydroxide-Mediated Nickel-Based Electrocatalysts for High-Current-Density Hydrogen Evolution in Alkaline Media

Yuting Luo[1†], Zhiyuan Zhang[1†], Fengning Yang[1], Jiong Li[2], Zhibo Liu[3], Wencai Ren[3], Shuo Zhang[2*] & Bilu Liu[1*]

1. Shenzhen Geim Graphene Center, Tsinghua-Berkeley Shenzhen Institute & Tsinghua Shenzhen International Graduate School, Tsinghua University, Shenzhen 518055, P. R. China.

2. Shanghai Synchrotron Radiation Facility, Shanghai Advanced Research Institute, Chinese Academy of Sciences, Shanghai 201210, China.

3. Shenyang National Laboratory for Materials Sciences, Institute of Metal Research, Chinese Academy of Sciences, Shenyang, Liaoning 110016, P. R. China.

† These two authors contributed equally.

* E-mail: bilu.liu@sz.tsinghua.edu.cn; zhangshuo@sinap.ac.cn

**Abstract**

Large-scale production of hydrogen by electrochemical water splitting is considered as a promising technology to address critical energy challenges caused by the extensive use of fossil fuels. Although nonprecious nickel-based catalysts work well at low current densities, they need large overpotentials at high current densities that hinders their potential applications in practical industry. Here we report a hydroxide-mediated nickel-based electrocatalyst for high-current-density hydrogen evolution, which delivers a current density of 1000 mA cm$^{-2}$ at an overpotential of 98 mV. Combined X-ray absorption spectroscopy and high-resolution X-ray photoelectron spectroscopy results show charge redistribution of nickel sites caused by Mo and surface FeO$_x$ clusters, which can stabilize the surface nickel hydroxide at high current densities for promoting water dissociation step. Such catalyst is synthesized at the metre scale and shows a current density of 500 mA cm$^{-2}$ at 1.56 V in the overall water splitting, which demonstrate its potential for practical use. This work highlights a charge-engineering strategy for rational design of catalysts that work well at high current densities.

**Introduction**

Hydrogen (H$_2$) is of great importance in key industrial processes, such as oil refining and production of ammonia.[1] It is clean and energy-dense, so that it can help to tackle critical energy challenges caused by the extensive use of fossil fuels.[2] Electrochemical water splitting that converts water into hydrogen and oxygen is a promising way for sustainable production of hydrogen, especially when it is driven by electricity generated by sunlight, wind, and other renewable energy.[3] It can be done in either acidic or alkaline media. The alkaline water splitting is used in industrial plants because it is cost-effective, long lifetime, and less corrosive to electrolyzer construction.[4, 5] Unlike to acidic water splitting, transitional metals, such as nickel (Ni), cobalt (Co), and iron (Fe), are stable in alkaline media. The prices of these metals are about five orders of magnitude lower than that of platinum (Pt),[6] which stir the passions of researchers to find transitional metal-based electrocatalysts for alkaline water splitting. Nevertheless, hydrogen evolution reaction (HER) in alkaline media is challenging because its kinetics is slowed by an additional water dissociation step.[7] The reaction rate on Pt, the best HER electrocatalyst, is usually

2-3 orders lower in alkali than that in the acid.[8]

Transitional metal-based electrocatalysts with high activity for alkaline HER have been explored over the past decade, including metal oxides,[9-11] metal alloys,[12-15] traditional metal phosphides,[16-18] carbides,[19, 20] and sulfides,[21-23] *etc*. Among them, Ni-based catalysts are promising for alkaline HER because of their high catalytic activity, high electrical conductivity, and low price. For example, Gong *et al*. reported that NiO/Ni nanoparticles anchored on the sidewalls of carbon nanotubes showed good HER performance and achieved ~20 mA cm$^{-2}$ at voltage of 1.5 V.[24] Nickel hydroxide on the catalyst surface could facilitate water dissociation step and provide protons for the following generation of $H_2$, rendering high activity of metal catalysts for alkaline HER.[21, 24] Alloying Ni with another metal (Ni-M, M = Zn, Al, Mo, Cr or Fe) could also improved its HER activity.[25, 26] Although these nickel-based catalysts operate well at low current densities, their performance at high current densities needs to be improved at perspectives of low overpotentials and excellent stabilities. The required current densities are from 200 to 500 mA cm$^{-2}$ at potentials of 1.8 to 2.5 V in the commercial alkaline electrolyzers.[4, 27] Raney Ni is the currently-used HER catalyst in alkaline water splitting that has advantages of low cost, large surface area, and decent stability, which however, delivers a current density of 500 mA cm$^{-2}$ at fairly large overpotentials of 300–500 mV.[28] The increased current density is desired but restricted by the large overpotentials.[29] It remains difficult for Ni-based catalysts that show good HER performance at high current densities.

Here, we report a Ni-based catalyst that is composed of hydroxide-mediated $Ni_4Mo$ nanoparticles decorated by $FeO_x$ and anchored on $MoO_2$ nanosheets (h-NiMoFe), which shows superior performance for alkaline HER at high current density. Combined X-ray absorption spectroscopy (XAS) and X-ray photoelectron spectroscopy (XPS) results reveal that the charge redistribution of nickel site is caused by both Mo and surface $FeO_x$ clusters, which stabilize the surface nickel hydroxide at high current densities and benefit the water dissociation step. As a result, the h-NiMoFe catalyst shows good performance for high-current-density HER with a Tafel slope of 32.7 mV dec$^{-1}$ and an overpotential of 62 mV at 500 mA cm$^{-2}$, especially a high current density of 1000 mA cm$^{-2}$ at 98 mV. Such a catalyst is synthesized at the metre scale with a price of ~82 US\$ m$^{-2}$. It is then used for the overall water splitting, which delivers 500 mA cm$^{-2}$ at 1.56 V and keeps stable over 40 h. Our results show potential of h-NiMoFe catalyst in practical use. This work may guide the rational design of electrocatalysts that work well at high current densities by a charge engineering strategy.

**Results and discussion**

The h-NiMoFe was prepared by a two-step method. Microspheres composed of radially aligned iron-doped nickel molybdate nanosheets (Fe-NiMoO$_4$) were first grown on Ni foams by the hydrothermal method, followed by reduction at 500 ºC in Ar/H$_2$ mixture flow (Fig. 1a-c, see details in the "Methods" section). The Fe-NiMoO$_4$ microspheres show a three-dimensional rugged feature under the scanning electronic microscope (SEM), which endow the sample a strong capillary force to promote electrolytes pumping onto this surface[30] and increase its affinity to electrolytes (Fig. 1e), in contrast to the Ni foam (Fig. 1d). Such microsphere structure maintains during H$_2$ reduction process and nanopores are formed on the basal planes of nanosheets due to partial removal of O during reduction (Fig. 1e). The h-NiMoFe shows an enhanced affinity to alkaline electrolytes and its mass transfer ability surpasses that of the Fe-NiMoO$_4$ (Fig. 1f), which may be caused by the existence of nanopores. These results show that h-NiMoFe with rugged morphology is obtained by a two-step approach, guaranteeing accessibility of the catalyst to alkaline electrolytes.

To shed light on the microscopic structure of the h-NiMoFe catalyst, spectroscopic characterization was carried out. High-resolution transmission electron microscope (HRTEM) images (Fig. 1g) show that the catalyst is composed of ultrathin nanosheets with nanoparticles anchoring on them (Fig. 1h). There are a large number of nanopores on the nanosheets, showing a narrow diameter distribution of 3.1 ± 1.2 nm. Profiles of the nanoparticles are distinct under high-angle annular dark-field scanning TEM (HAADF-STEM), showing an average diameter of 18.0 ± 5.7 nm (Fig. 1i). Whereas Mo and O elements are distributed uniformly in nanosheets at energy dispersive X-ray spectroscopy (EDS) elemental maps, Ni and Fe elements mainly exist in the nanoparticles (Figs. 1i, S1 and S2), revealing different chemical compositions of nanosheets and the nanoparticles. The HRTEM images (Fig. 1j) show typical lattice spacings of 0.24 nm and 0.20 nm, corresponding to the (020) plane of $MoO_2$ and (220) plane of $Ni_4Mo$, respectively. In combination with the EDS results, nanoparticles are made up of $Ni_4Mo$ and nanosheets are $MoO_2$. The $Ni_4Mo$ nanoparticles are anchored on the $MoO_2$ nanosheets (Fig. 1k). The compositions of $Ni_4Mo$ and $MoO_2$ in h-NiMoFe catalyst are further confirmed by the X-ray diffraction (XRD) (Fig. S3) and Raman spectrum (Fig. S4). $MoO_2$ nanosheets are frameworks of the microspheres and work as high-surface-area and electroconductive supports.[31, 32] Although Fe element is observed under EDS, no Fe-based compounds are detected by HRTEM, XRD, and Raman, indicating the lack of long-range crystal structure. There results show that h-NiMoFe is composed of Fe-decorated $Ni_4Mo$ nanoparticles that are anchored on $MoO_2$ supports.

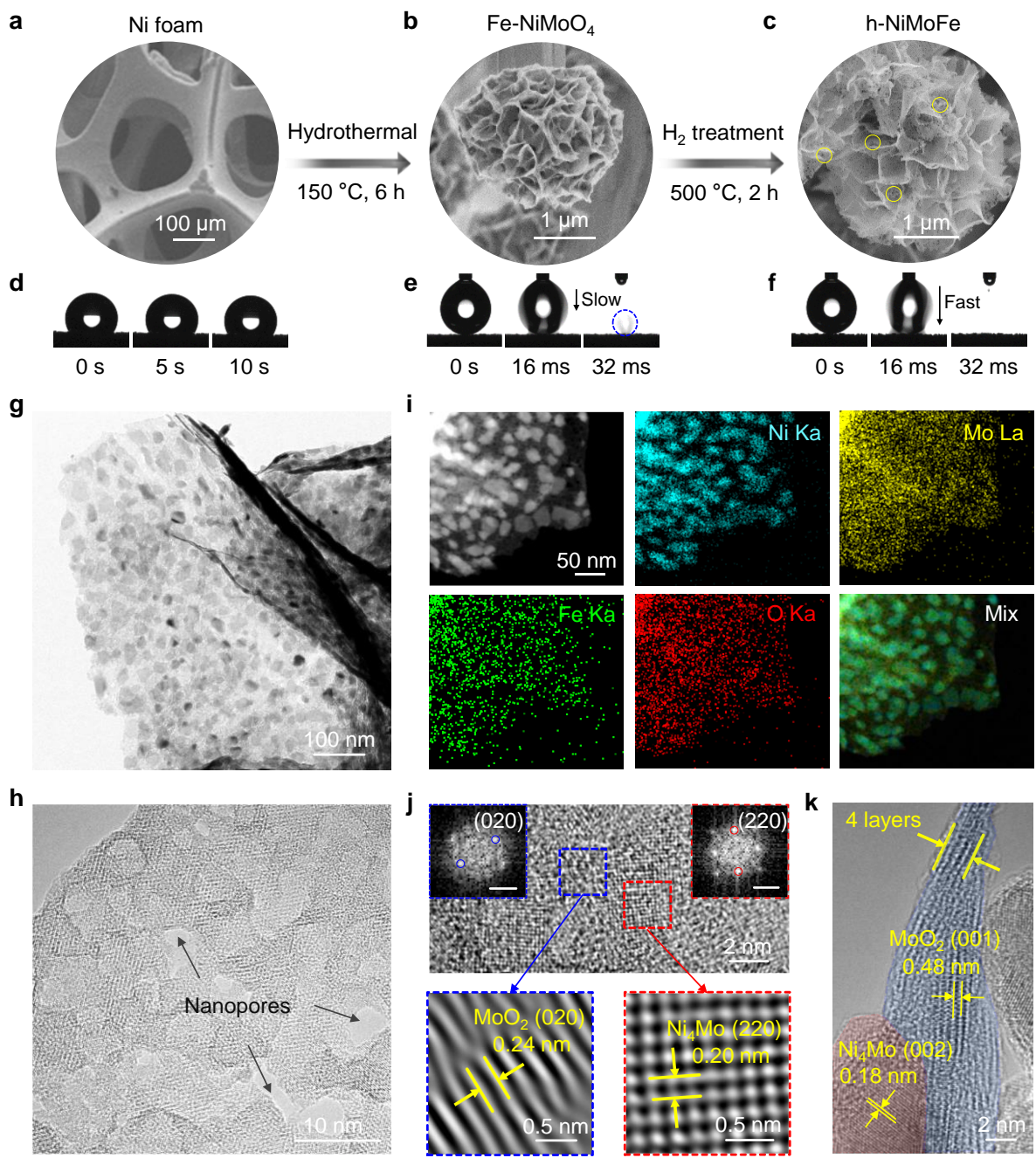

**Figure 1. Synthesis and characterization of h-NiMoFe catalyst.** Synthesis of the catalyst by a two-step process from (a) bare Ni foams via (b) oxide precursors to (c) h-NiMoFe. (d-f) Captured photos showing the dynamic wettability on different samples. The droplets are 1.0 M KOH with an identical volume of 4 μL. (g, h) HRTEM images of the catalyst, where small nanoparticles are anchored on porous nanosheets. (i) EDS mappings of the catalysts showing that Mo and O elements are uniformly

distributed while Ni and Fe elements mainly exist on the nanoparticles. HRTEM images of h-NiMoFe, showing (j) an enlarged view of h-NiMoFe, lattice fringes of $MoO_2$ nanosheet (blue box) and $Ni_4Mo$ nanoparticles (red box), as well as (k) a side view of the $MoO_2$ nanosheet.

More insights into the local electronic structure and local geometric structure of Fe and Mo in h-NiMoFe catalyst are obtained from X-ray absorption fine structure (XAFS) spectroscopy at the Fe K-edges and Mo K-edge. Figure 2a displays the Fe K-edge X-ray absorption near-edge structure (XANES) for h-NiMoFe and reference samples. The position of adsorption edge indicates that the oxidation state of Fe in h-NiMoFe is in between +2.5 and +3, suggesting that Fe are not incorporated into $Ni_4Mo$ lattice. Consistent results are seen in the Fournier transforms of the extended X-ray absorption fine structure (FT-EXAFS) at Fe K-edge (Fig. 2b, Fig. S5 and Table S1). The FT-EXAFS spectrum of h-NiMoFe shows a scattering peak at ~1.5 Å, corresponding to the Fe−O distance. Nevertheless, the absence of Fe-metal scattering peak shows that Fe species lack long-range crystal structure, agreeing well with the XRD result. The further fitting results unravel a four-fold coordination environment with a Fe−O bond length of 1.92 Å. These data suggest that irons are on $Ni_4Mo$ surface in oxidation, which are possibly in form of mononuclear $FeO_x$ clusters. With respect to molybdenum, mixture of Mo oxide and alloy in h-NiMoFe is shown in Mo K-edge FT-EXAFS spectrum, as evidenced by the two peaks at ~1.7Å and ~2.6 Å (Fig. S6), attributable to Mo−O and Mo−M (M = Ni or Mo) bonds in $MoO_2$ and $Ni_4Mo$, respectively. The above results confirm that iron shows oxidation nature and molybdenum the alloy nature in $Ni_4Mo$ nanoparticles in the designed catalysts.

The effects of presence of Mo/Fe on local electronic structure of Ni were further investigated by the Ni K-edge XAS (Fig. 2c and 2d). The FT-EXAFS spectra of h-NiMoFe, NiMo, and Ni foil show that nickel is in metallic state among the three samples, as evidenced by a scatter peak at ~2.2 Å that is attributable to Ni-M (M = Ni or Mo) bond (Fig. 2d, Fig. S7 and Table S2). The introduction of Mo/Fe remarkably change the charge distribution of nickel sites, as revealed by the Ni K-edge XANES (Fig. 2c). Compared to Ni foil, the absorption edge in NiMo sample moves to higher energy and the white-line intensity increases (Fig. 2c), showing that nickel in NiMo sample has more unoccupied state than that in Ni foil due to the presence of Mo. This effect is further enhanced under the presence of Fe. These results indicate a charge transfer from Ni to Mo/Fe in h-NiMoFe. The XPS characterization also unraveled that the peak of $Ni^0$ shifts to higher binding energy when Mo/Fe exists (Fig. S8), in consistent with XAS results. These comprehensive characterizations demonstrate a strong interaction between Ni and Mo/Fe in h-NiMoFe catalyst and modulate the charge distribution of nickel sites.

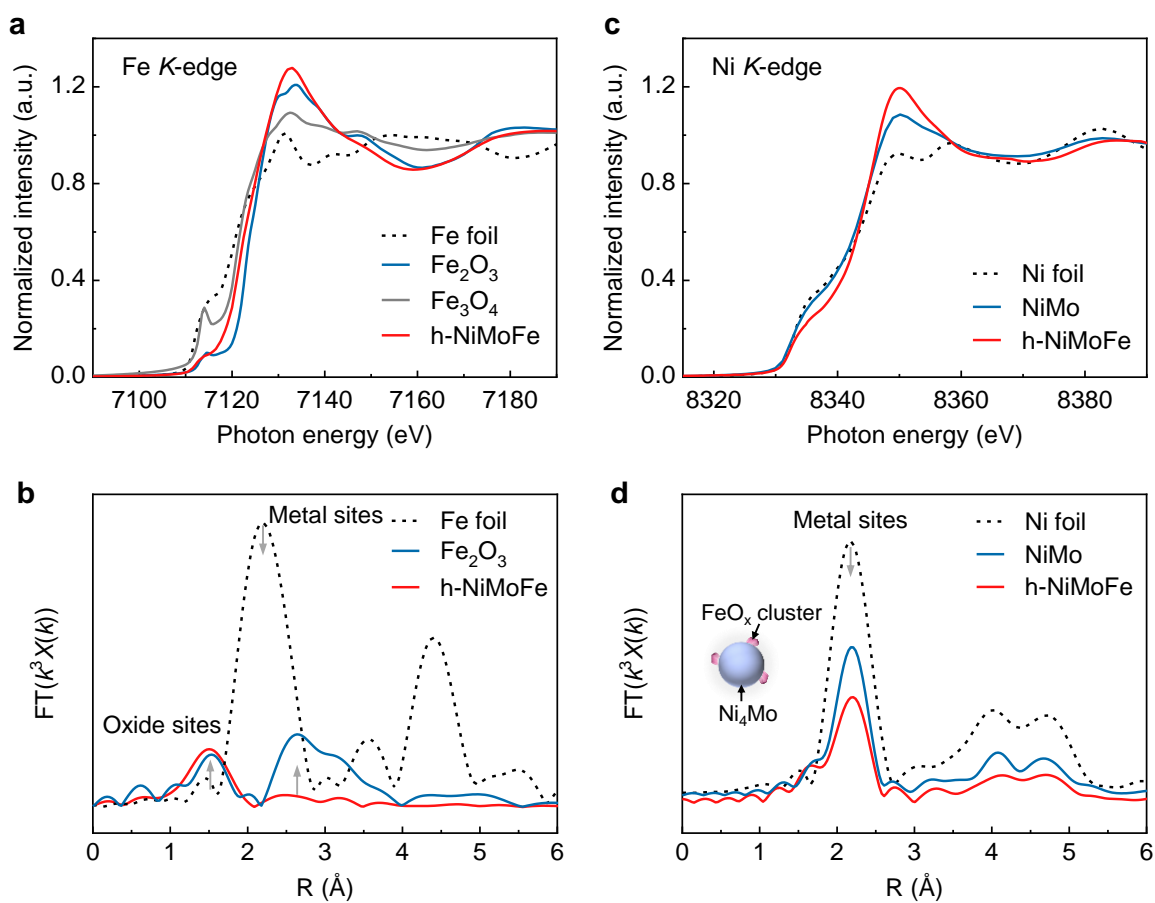

**Figure 2. X-ray absorption spectroscopies.** (a, c) X-ray absorption and (b, d) Fourier analysis of the EXAFS of h-NiMoFe catalyst and controls at Fe and Ni K-edges in absorption mode, showing the effect of Mo and Fe on the electronic structure and local environment of Ni.

To understand the impacts of presentence of Mo/Fe in h-NiMoFe on its surface nickel species, we used high-resolution XPS characterization. It reveals that nickel exists as $Ni(OH)_2$, NiO, and metallic Ni in curve-fitting of the Ni 3$s$ XPS spectra for h-NiMoFe, NiMo, and Ni samples (Fig. 3a and Fig. S9). We observed a higher content of surface hydroxide on h-NiMoFe than those on control NiMo and Ni samples (Fig. 3b). Specifically, more than 60% of nickel is in form of $Ni(OH)_2$ on h-NiMoFe surface, higher than the contents of surface metallic Ni and NiO. This trend is distinct from that shown by control Ni sample, where metallic Ni is dominated on the surface. It is also different to NiMo sample, where the metallic Ni is prone to convert to surface NiO rather than $Ni(OH)_2$. These results indicate that the existence of Mo/Fe may induce the conversion of Ni surface to $Ni(OH)_2$ because nickel in h-NiMoFe has more unoccupied state than those in control NiMo and Ni samples. The h-NiMoFe catalyst has a hydroxide-mediated surface. We then compared the Ni 3$s$ XPS spectra of h-NiMoFe, NiMo, and Ni samples before and after HER cycling at potential windows corresponding to their current densities from 0 to 1000 mA cm$^{-2}$. All samples were carefully treated to ensure that they were exposed to the ambient environment for less than 1 min (see details in the "Methods" section). We observed that the metallic Ni still dominates on Ni sample after HER (Fig. 3b and Fig. S10). This phenomenon is distinct from that exhibited by NiMo sample, where most of surface metallic Ni are converted to $Ni(OH)_2$ (Fig. 3b and Fig. S10). With respect to h-NiMoFe sample, the conversion from surface metallic Ni to $Ni(OH)_2$ is similar and the content of $Ni(OH)_2$ is even higher, in this way enhancing the water dissociation of the Ni surface.[33] As a result, the h-NiMoFe catalysts show a better HER activity than the Ni and NiMo catalysts (Figs. S11 and S12). These data show that h-NiMoFe has the ability to stabilize the surface hydroxide during HER process at high current densities.

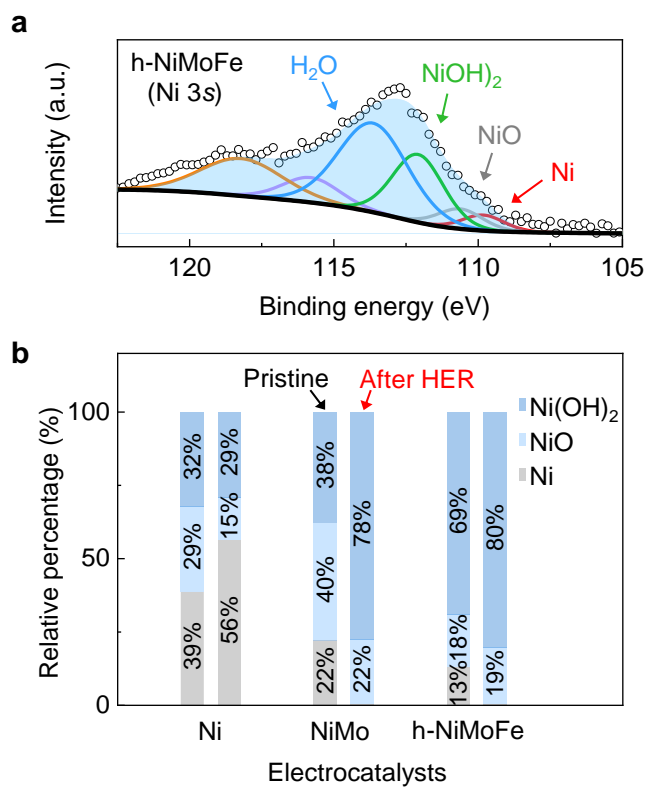

**Figure 3. Surface chemistry characterization before and after HER.** (**a**) A XPS Ni 3$s$ spectrum of h-NiMoFe showing its surface compositions. (**b**) Relative percentages of surface Ni species on Ni, NiMo, and h-NiMoFe samples before and after experiencing HER cycles from 0 to 1000 mA cm$^{-2}$. These results are obtained from their XPS Ni 3$s$ spectra.

The HER performance of h-NiMoFe at high current densities was characterized using a three-electrode electrochemical cell in a 1.0 M KOH solution. Figure 4a shows the polarization curves of h-NiMoFe, Ni foam, and Pt catalysts. The onset overpotential of h-NiMoFe is closed to 0 mV, which is comparable to that of Pt catalyst and smaller than that of Ni foam (~90 mV), indicating the high HER activity of h-NiMoFe. The consistent results are shown by their Tafel curves, where h-NiMoFe has a Tafel slope of 32.7 mV dec$^{-1}$ that is also closed to Pt (Fig. 4b). The h-NiMoFe shows the smallest charge transfer resistance among these samples (Fig. S13). In the case of low current densities, h-NiMoFe is similar to Pt and surpasses Ni foam. For example, the overpotentials of h-NiMoFe, Pt and Ni foam 10 mA cm$^{-2}$ are 10 mV, 59 mV, and 309 mV, respectively (Fig. 4a). In the case of high current densities, h-NiMoFe needs smaller overpotentials to deliver the same current densities compared to Pt and Ni foam. For instance, the overpotential is 62 mV at 500 mA cm$^{-2}$ for h-NiMoFe, while Pt needs 402 mV to deliver the same current density (Fig. 4a). Moreover, h-NiMoFe shows an increased current density of 1000 mA cm$^{-2}$ at an overpotential of 98 mV. These results show that h-NiMoFe catalyst has superior performance for high-current-density HER.

To further evaluate the high-current-density HER performance of h-NiMoFe catalyst, we analyzed the relationships between current densities and $\Delta\eta/\Delta\log|j|$ (Fig. 4c). Specifically, $\Delta\eta/\Delta\log|j|$ ratio shows how much overpotential ($\eta$) is needed when current density ($j$) increases, which is maintained small on good catalysts. The ratio for h-NiMoFe remains small (~23 mV dec$^{-1}$), but those for Pt increase to more than 40 mV dec$^{-1}$ when current density increases to 50 mA/cm$^2$. As current density is even larger (*e.g.*, 500 mA cm$^{-2}$), the ratio for h-NiMoFe is still much smaller than ~284 mV dec$^{-1}$ of Pt catalyst. Moreover, it is smaller than 120 mV dec$^{-1}$ for h-NiMoFe even the current density is high up to 1000 mA cm$^{-2}$. These results further confirm that the h-NiMoFe catalyst works well for high-current-density HER. The advantage of h-NiMoFe to deliver high current densities at small overpotentials is seen by comparation with the previously reported data (Fig. 4d and Table S3). The h-NiMoFe shows a current density of 1026 mA cm$^{-2}$ at 100 mV. By comparing the HER performance of h-NiMoFe with those of the control NiMo and Ni catalysts, it is suggested that alloying Ni$_4$Mo as well as decoration of FeO$_x$ are paly vital roles in the good performance of h-NiMoFe catalyst (Figs. S11 and S12). Taking together, the designed catalyst shows high current densities at small overpotentials, further confirming its good performance for HER.

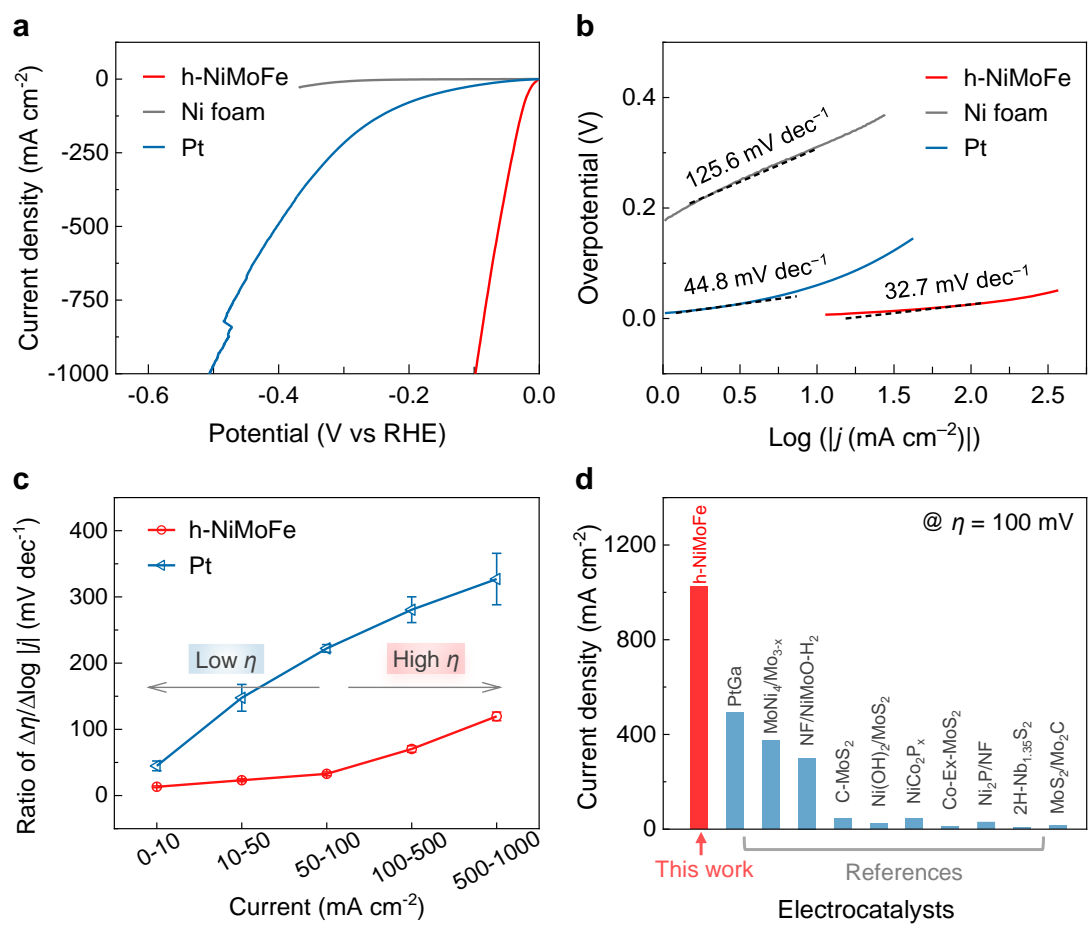

**Figure 4. HER performance at high current densities.** (**a**) Polarization curves and (**b**) Tafel plots of h-NiMoFe catalyst and its controls. (**c**) Ratios of $\Delta\eta/\Delta\log|j|$ of the catalysts at different current densities. (**d**) A comparison of current densities at an overpotential of 100 mV, which are achieved on different electrocatalysts including the h-NiMoFe catalyst and other catalysts reported in previous literatures.

To realize the large-scale use of electrolyzers, it is important to reduce the cost of device components. Scaling-up synthesis of catalysts from low-cost precursors can pave the way for this goal. To show the feasibility of scaling-up synthesis of h-NiMoFe catalyst, we used a roll of Ni foam with a size of 1.5 meter × 0.1 meter as the substrate and then grew h-NiMoFe on it by the same synthesis method (Fig. 5a, see details in the "Methods" section). The price of such catalyst is ~82 US$ m$^{-2}$ (Table S4 and Note S1), which is close to commercial Raney Ni (~60 US$ m$^{-2}$) and ~50 times lower than commercial Pt/C catalyst on carbon paper (Table S5). Such a low price is attributed to cheap precursors with transitional metal elements (Ni, Fe, and Mo) having higher global availability than Pt.[34] Moreover, the high Earth reserves of these metals promise the potential of mass production of h-NiMoFe catalyst. Together, the h-NiMoFe catalyst shows not only high performance but also low price.

For practical applications, Raney nickel catalysts require operation for overall water splitting in the same alkaline electrolyte and deliver high current densities. We characterize the electrochemical performance of h-NiMoFe catalyst in overall water splitting systems (Fig. 4b). This catalyst is of high performance and achieves a current density of 500 mA cm$^{-2}$ at the record-low cell potential of 1.56 V (Fig. 4c). It outperforms the Raney nickel catalyst under these conditions or even favorable conditions, which requires cell potentials of 1.8 to 2.5 V to achieve 200 to 500 mA cm$^{-2}$.[27] Owing to such good performance, the overall water splitting system using h-NiMoFe catalysts can be powered by a mini solar photovoltaic cell whose outputs voltage is smaller than 2 V (Fig. 4d). Furthermore, the Faradaic efficiency (FE) is close to 100% by measuring the volumes of hydrogen generation (Fig. 4e). Stability is another parameter of catalysts, especially under current densities relevant to industry. The h-NiMoFe catalyst shows no obvious degradation of performance at various current densities over a period of 40 h continuous operation (Fig. 4d and Fig. S14), confirming its good stability. The advantage of this catalyst in terms of delivering high current densities for water splitting is seen by comparing it with previously reported catalysts (Fig. 4e and Table S6). These results confirm a strong potential of the h-NiMoFe catalyst for practical use.

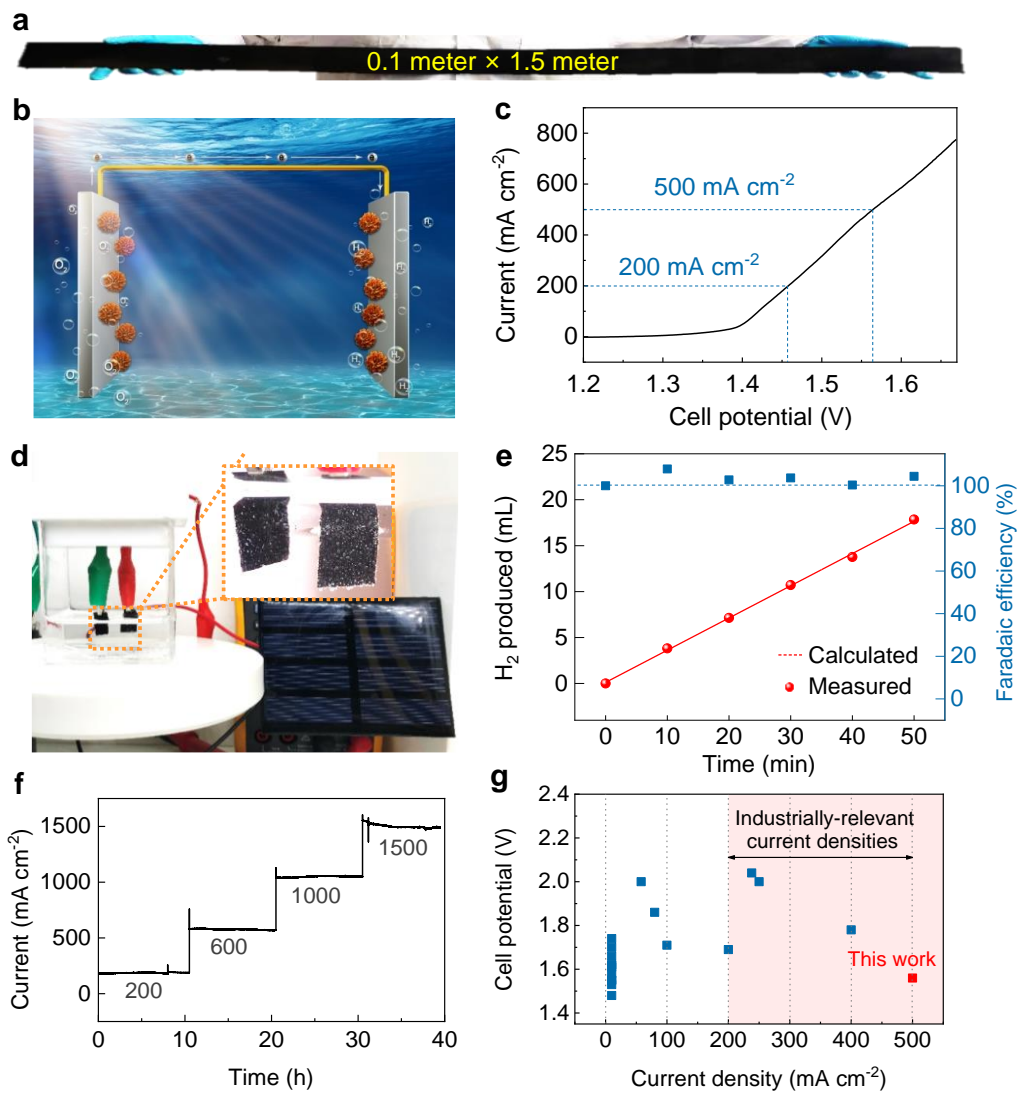

**Figure 5. Scaling-up synthesis of h-NiMoFe catalyst and the overall water splitting.** (**a**) A scheme showing the overall water splitting by h-NiMoFe catalyst. (**b**) Scaling-up synthesis of h-NiMoFe with a size of 0.1 meter × 1.5 meter. (**c**) A demo showing the overall water splitting driven by a commercial photovoltaic cell. (**d**) Polarization curve and (**e**) time-current curve showing the catalytic performance of h-NiMoFe for the overall water splitting. (**f**) A comparation of overall water splitting performance of the h-NiMoFe catalysts with previously reported data.

**Conclusions**

We have developed a brand-new h-NiMoFe electrocatalyst that is highly active and stable, performing especially well at high current densities up to 1000 mA cm$^{-2}$. Detailed microstructure characterization shows that strong interactions between Ni and Mo/Fe modulate the local electronic structure of Ni with more hydroxide on the surface than control samples. As a result, h-NiMoFe catalyst could stabilize the surface hydroxide on it, even at high current densities. Such catalyst could be synthesized in a meter level and be used for overall water splitting, delivering a current density of 500 mA cm$^{-2}$ at 1.56 V and a stability over 40 h. These findings shine lights on charge-engineering strategy for rational design of electrocatalysts that work well at high current densities.

**Supporting Information**

Supplementary data associated with this article can be found in the online version.

**Acknowledgements**

We sincerely thank Prof. Hui-Ming Cheng for fruitful discussions and inputs to this work. We acknowledge support from the National Natural Science Foundation of China (Nos. 51722206 and 51920105002), the Youth 1000-Talent Program of China, Guangdong Innovative and Entrepreneurial Research Team Program (No. 2017ZT07C341), the Bureau of Industry and Information Technology of Shenzhen for the "2017 Graphene Manufacturing Innovation Center Project" (No. 201901171523). We also thank stuffs in BL11B beamline in Shanghai Synchrotron Radiation Facility (SSRF) for their technical assistance.

**Author contributions**

Y.L. and B.L. conceived the idea. Y.L., L.T., and F.Y. prepared materials. Y.T. performed XRD, SEM, Raman, and XPS characterization and electrochemical experiments. Z.L. and W.C. performed TEM characterization. F.Y. took part in the electrochemical measurements and discussion. S.Z., Z.Z, L.J. and Y.L. performed and analyzed XAS results. B.L. supervised the project and directed the research. Y.L., S.Z. and B.L. interpreted the results. Y.L., S.Z. and B.L. wrote the manuscript with feedbacks from the other authors.

**Declaration of interests**

The authors declare no competing interests.

**TOC**

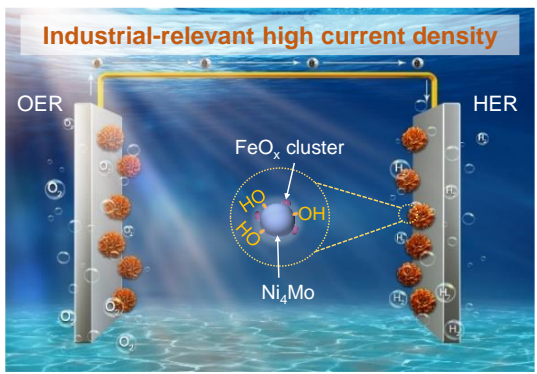